\documentclass[preprint,aps,12pt,preprintnumbers,eqsecnum,nofootinbib]{revtex4}
\usepackage{graphicx, color}
\usepackage{amssymb,amsmath}
\def\lag{{\mathcal{L}}}

\def\be{\begin{equation}}
\def\ee{\end{equation}}
\def\beq{\begin{eqnarray}}
\def\eeq{\end{eqnarray}}
\def\f{\frac}
\def\lf{\left}
\def\rh{\right}
\begin{document}
%
%
\title{The Galactic Center Region Gamma Ray Excess from \\ A Supersymmetric Leptophilic Higgs Model}
\author{Gardner Marshall}\email[]{grmarshall@email.wm.edu}
\author{Reinard Primulando}\email[]{rprimulando@email.wm.edu}
\affiliation{Particle Theory Group, Department of Physics,
College of William and Mary, Williamsburg, VA 23187-8795}
\date{February 2011}
\begin{abstract}
In a recent paper by Hooper and Goodenough, data from the Fermi
Gamma Ray Telescope was analyzed and an excess of gamma rays was
claimed to be found in the emission spectrum from the Galactic
Center Region. Hooper and Goodenough suggest that the claimed excess
can be well explained by 7-10 GeV annihilating dark matter with a
power law density profile if the dark matter annihilates
predominantly to tau pairs. In this paper we present such a dark
matter model by extending the MSSM to include four Higgs doublets
and one scalar singlet. A $\mathbb{Z}_{2}$ symmetry is imposed that
enforces a Yukawa structure so that the up quarks, down quarks, and
leptons each receive mass from a distinct doublet. This leads to an
enhanced coupling of scalars to leptons and allows the model to
naturally achieve the required phenomenology in order to explain the
gamma ray excess. Our model yields the correct dark matter thermal
relic density and avoids collider bounds from measurements of the
$Z$ width as well as direct production at LEP.
\end{abstract}
\maketitle

\section{Introduction}
\label{sec:intro}

Recently, Hooper and Goodenough examined the first two years of
Fermi Gamma Ray Space Telescope (FGST) data from the inner
$10^\circ$ around the Galactic Center \cite{Hooper:2010mq}. They
found that the gamma ray emissions coming from between $1.25^\circ$
and $10^\circ$ of the Galactic Center is consistent with what is
expected from known emission mechanisms such as cosmic rays
colliding with gas to produce subsequently decaying pions, inverse
Compton scattering of cosmic ray electrons, and known gamma ray
point sources. {In order to model the gamma ray
background within $2^{\circ}$ of the Galactic Center, Hooper and
Goodenough model the emission of the Galactic black hole Sgr A* as a
power-law extrapolated from higher energy HESS observations.
Comparing the FGST measurements to this background, Hooper and
Goodenough found that it agrees very well with FGST data between
$1.25^{\circ} - 2^{\circ}$ but found an excess in the observed gamma
ray intensity within $1.25^{\circ}$. It has been pointed out by Ref.
\cite{Boyarsky:2010dr} however, that a simple power-law
extrapolation of HESS data may understate the flux of the central
point source Sgr A* as the slope of its spectrum may deviate from
the constant HESS results below an energy of ~100 GeV.}

The authors of Ref. \cite{Hooper:2010mq} showed that the increased
gamma ray emissions are well described by annihilating dark matter
that has a cusped halo profile ($\rho \propto r^{-\gamma}$, with
$\gamma = 1.18$ to $1.33$) provided that the dark matter satisfies
three basic conditions. The conditions required of the dark matter
are 1) that it have a mass between $7-10$ GeV, 2) that it annihilate
into $\tau$-pairs most of the time, but into hadronic channels
$15-40 \%$ of the time, and 3) that its total annihilation cross
section yield a thermal average within the range $\langle \sigma v
\rangle = 4.6 \times 10^{-27} - 5.3 \times 10^{-26} \ \textrm{
cm}^{3}/\textrm{s}$. {It should be noted that the
results of Hooper and Goodenough are controversial, and the
Fermi-LAT collaboration itself has not yet published official
results. In addition, other background related explanations for the
gamma ray excess have been proposed such as the existence of a
pulsar near the Galactic Center \cite{Abazajian:2010zy}. In this
paper we proceed with the assumption that the analysis of Hooper and Goodenough is correct.
The astrophysical and particle physics implications of this finding
are discussed in Refs. \cite{Kopp:2010su, Kelso:2010sj}.}

In this paper we construct a dark matter model satisfying the above
conditions by adding a singlet to the supersymmetric leptophilic
Higgs model (SLHM) \cite{Marshall:2010qi}. In the SLHM the up
quarks, down quarks, and leptons, each receive mass from a separate
Higgs doublet. For our purposes, the salient characteristic of the
SLHM is that it endows the leptons with an enhanced coupling to one
of the scalars. This provides a natural mechanism for dark matter
particles to annihilate predominantly into $\tau$-pairs. This model
of dark matter is able to successfully account for the FGST
observations, yields the correct relic density, and evades relevant
collider bounds such as measurements of the $Z$ width and direct
production at LEP. {The idea of a leptophilic Higgs
has been studied as a possible explanation for the $e^{\pm}$ excess
observed by PAMELA and ATIC in Ref. \cite{Goh:2009wg}. However, this
entails a 100 GeV - 1 TeV dark matter particle, while our model
requires a light, $\mathcal{O}(10)$ GeV dark matter particle.} There
also exist some other models that can explain the Galactic Center
gamma ray excess \cite{Logan:2010nw}.

In addition to explaining the FGST observations, such a model of
light dark matter is also capable of describing observations by the
CoGeNT \cite{Aalseth:2010vx} and DAMA collaborations
\cite{Bernabei:2008yi}. CoGeNT has recently reported direct
detection signals that hint at the presence of $\mathcal{O}(10)$ GeV
dark matter compatible with the light dark matter interpretation of
DAMA's annual event rate modulation. Ref. \cite{Hooper:2010uy}
showed that dark matter with a mass between $7-8$ GeV that has a
spin independent cross section approximately between $ \sigma_{SI} =
1 \times 10^{-40} - 3 \times 10^{-40} \textrm{ cm}^2$ is consistent
with both CoGeNT and DAMA signals. Although the XENON
\cite{Aprile:2010um} and CDMS \cite{Ahmed:2010wy} collaborations
challenge this report, Ref. \cite{Kelso:2010sj} has pointed out that
``zero-charge" background events lie in the signal region. The
authors suggest that the bound could possibly be loosened if a
modest uncertainty or systematic error is introduced in the energy
scale calibration near the energy threshold. Although our model is
able to explain the reported observations of the CoGeNT and DAMA
collaborations, it is not dependent upon their validity. By simply
moving to another region of parameter space our model can coexist
with the absolute refutation of CoGeNT and DAMA while continuing to
explain the FGST results and avoiding collider bounds.

Our paper is organized as follows. In Section \ref{sec:model} we
introduce the setup of the model and calculate the mass matrices for
the scalars and the neutralinos. In Section \ref{sec:annihilation}
we describe the process by which the dark matter annihilates into
Standard model particles and calculate the relevant cross sections
for a benchmark point in parameter space. We also show that the
resultant relic density is consistent with current cosmological
measurements. In Section \ref{sec:direct} we discuss possible direct
detection and in Section \ref{sec:bounds} we discuss relevant bounds
for this model and show that it is currently viable. Lastly, we
conclude with Section \ref{sec:conc} and summarize the results of
the paper.

\section{The Model}
\label{sec:model}

In this model the quark and lepton content is that of the MSSM. To
this we add four Higgs doublets, $\widehat H_{u}$, $\widehat H_{d}$,
$\widehat H_{0}$, and $\widehat H_{\ell}$, with weak hypercharge
assignment $+1/2$, $-1/2$, $+1/2$, and $-1/2$ respectively.
{The third Higgs doublet is necessary to achieve a
leptonic structure, while the fourth doublet is required for anomaly
cancelation. In order to avoid problems with the $Z$ decay width, we
introduce a singlet $\widehat S$ that acts as $\mathcal{O} (10)$ GeV
dark matter. The idea of adding a light singlet to the MSSM to act
as dark matter was also considered in \cite{Kappl:2010qx}, while the
use of a singlet for other purposes such as solving the $\mu$
problem was first developed in \cite{quiesp}.} The superpotential is
given by \begin{equation} \label{eq:superpotential} \begin{split} W
= {} & y_{u} \widehat{U}\widehat{Q}\widehat{H}_{u} - y_{d}
\widehat{D}\widehat{Q}\widehat{H}_{d} - y_{\ell}
\widehat{E}\widehat{L}\widehat{H}_{\ell} + \mu_{q} \widehat{H}_{u}
\widehat{H}_{d} + \mu_{\ell} \widehat{H}_{0} \widehat{H}_{\ell} \\ &
+ \kappa_{q} \widehat{S} \widehat{H}_{u} \widehat{H}_{d} +
\kappa_{\ell} \widehat{S} \widehat{H}_{0} \widehat{H}_{\ell} +
\lambda_{1}^{2} \widehat{S} + \frac{1}{\hspace{.02 in} 2}
\hspace{.03 in} \lambda_{2} \widehat{S}^{2} + \frac{1}{\hspace{.02
in} 3} \hspace{.03 in} \kappa_{s} \widehat{S}^{3}, \end{split}
\end{equation} where the hats denote superfields. In the
superpotential we introduced a $\mathbb{Z}_{2}$ symmetry under which
$\widehat{H}_{0}$, $\widehat{H}_{\ell}$ and $\widehat{E}$ are odd
while all other fields are even. The symmetry enforces a Yukawa
structure in which $\widehat{H}_{u}$ gives mass to up-type quarks,
$\widehat{H}_{d}$ to down-type quarks, and $\widehat{H}_{\ell}$ to
leptons, while $\widehat{H}_0$ does not couple to the quarks or
leptons and is called the inert doublet. It is introduced to ensure
anomaly cancellation. The $\mathbb{Z}_{2}$ symmetry is broken in
$V_{\textrm{soft}}$ so that we have: \footnote{In Ref.
\cite{Marshall:2010qi} the soft breaking terms
$m_{u0}^{2}H_{u}^{\dag}H_{0} + m_{d\ell}^{2}H_{d}^{\dag}H_{\ell} +
\textrm{h.c.}$ were omitted.} \begin{equation} \label{eq:vsoft}
\begin{split} V_{\textrm{soft}} = {} & m_{u}^{2} |H_{u}|^{2} +
m_{d}^{2} |H_{d}|^{2} + m_{0}^{2} |H_{0}|^{2} + m_{\ell}^{2}
|H_{\ell}|^{2} + m_{s}^{2} |S|^{2} \\ & + \Big(\mu_{1}^{2} H_{u}
H_{d} + \mu_{2}^{2} H_{0} H_{\ell} + \mu_{3}^{2} H_{u} H_{\ell} +
\mu_{4}^{2} H_{0} H_{d} \\ & + \mu_{a} S H_{u} H_{d} + \mu_{b} S
H_{0} H_{\ell} + \mu_{c} S H_{u} H_{\ell} + \mu_{d} S H_{0} H_{d} \\
& + m_{u0}^{2}H_{u}^{\dag}H_{0} + m_{d\ell}^{2}H_{d}^{\dag}H_{\ell}
+ t^{3} S + b_{s}^{2} S^{2} + a_{s} S^{3} + \textrm{h.c.} \Big).
\end{split}
\end{equation}

The breaking of the $\mathbb{Z}_{2}$ symmetry is discussed in
greater detail in Appendix \ref{sec:app2}. The Higgs sector
potential is given by $V = V_{D} + V_{F} + V_{\textrm{soft}}$.
Letting $\sigma^{a}$ denote the Pauli matrices for $a = 1,2,3$, the
D-term is simply \begin{equation}
\begin{split} V_{D} = {} & \frac{g^{2}}{8} \sum_{a} \left|
H_{u}^{\dag} \sigma^{a} H_{u} + H_{d}^{\dag} \sigma^{a} H_{d} +
H_{0}^{\dag} \sigma^{a} H_{0} + H_{\ell}^{\dag} \sigma^{a} H_{\ell}
\right|^{2} \\ & + \frac{g'^{ \hspace{.02 in} 2}}{8} \hspace{.02 in}
\Big| |H_{u}|^{2} - |H_{d}|^{2} + |H_{0}|^{2} - |H_{\ell}|^{2}
\Big|^{2}, \end{split} \end{equation} where $g$ and $g'$ are the
$SU(2)$ and $U(1)$ gauge couplings respectively. The F-term and
$V_{\textrm{soft}}$ combine with the D-term to yield the following
potential \begin{equation} \label{eq:higgspotential} \begin{split} V
= {} & \big( \mu_{q}^{2} + m_{u}^{2} \big) |H_{u}|^{2} + \big(
\mu_{q}^{2} + m_{d}^{2} \big) |H_{d}|^{2} + \big( \mu_{\ell}^{2} +
m_{0}^{2} \big) |H_{0}|^{2} + \big( \mu_{\ell}^{2} + m_{\ell}^{2}
\big) |H_{\ell}|^{2} \\ & + \Big[\big(\mu_{1}^{2} + \kappa_{q}
\lambda_{1}^{2} \big) H_{u} H_{d} + \big(\mu_{2}^{2} + \kappa_{\ell}
\lambda_{1}^{2} \big) H_{0} H_{\ell} + \mu_{3}^{2}H_{u} H_{\ell} +
\mu_{4}^{2}H_{0} H_{d} + \textrm{h.c.} \Big] \\ & + \Big| \kappa_{q}
H_{u} H_{d} + \kappa_{\ell}H_{0}H_{\ell}\Big|^{2} + \Big(
m_{u0}^{2}H_{u}^{\dag}H_{0} + m_{d\ell}^{2}H_{d}^{\dag}H_{\ell} +
\textrm{h.c.} \Big) + \big( m_{s}^{2} + \lambda_{2}^{2} \big)
|S|^{2} \\ & + \Big[\big(t^{3} + \lambda_{1}^{2} \lambda_{2}\big)S +
\big(b_{s}^{2} + \kappa_{s} \lambda_{2}^{2} \big)S^{2} + a_{s} S^{3}
+ \textrm{h.c.}\Big] + \kappa_{s} \lambda_{2} |S|^{2}\big(S
+S^{*}\big) + \kappa_{s}^{2}|S|^{4} \\ & + \Big[ \mu_{a} \big(H_{u}
H_{d}\big)S + \mu_{b} \big(H_{0} H_{\ell}\big)S + \mu_{c} \big(H_{u}
H_{\ell} \big)S + \mu_{d} \big(H_{0} H_{d}\big)S +
\textrm{h.c.}\Big] \\ & + \Big\{ \lambda_{2} \Big[ \kappa_{q}
\big(H_{u} H_{d}\big) + \kappa_{\ell} \big(H_{0}
H_{\ell}\big)\Big]S^{*} + \kappa_{s} \Big[\kappa_{q} \big(H_{u}
H_{d}\big) + \kappa_{\ell} \big(H_{0} H_{\ell}\big)\Big](S^{2})^{*}
+ \textrm{h.c.}\Big\} \\ & + \Big\{ \kappa_{q} \mu_{q}
\Big(|H_{u}|^{2} + |H_{d}|^{2}\Big) + \kappa_{\ell} \mu_{\ell}
\Big(|H_{0}|^{2} + |H_{\ell}|^{2}\Big)\Big\} \big(S + S^{*}\big) \\
& + \kappa_{q}^{2} \Big(|H_{u}|^{2} + |H_{d}|^{2}\Big)|S|^{2} +
\kappa_{\ell}^{2} \Big( |H_{0}|^{2} + |H_{\ell}|^{2}\Big)|S|^{2} +
V_{D}.
\end{split} \end{equation}

The singlet $S$ acquires the vev $\langle S \rangle =
v_{s}/\sqrt{2}$ while the Higgs doublets acquire the vevs:
\begin{equation} \langle H_{u} \rangle = \frac{1}{\sqrt{2}} \left(
\begin{array}{c} 0 \\ v_{u} \\ \end{array} \right), \hspace{.1 in}
\langle H_{d} \rangle = \frac{1}{\sqrt{2}} \left( \begin{array}{c}
v_{d} \\ 0 \\ \end{array} \right), \langle H_{0} \rangle =
\frac{1}{\sqrt{2}} \left( \begin{array}{c} 0 \\ v_{0} \\ \end{array}
\right), \hspace{.1 in} \langle H_{\ell} \rangle =
\frac{1}{\sqrt{2}} \left( \begin{array}{c} v_{\ell} \\ 0 \\
\end{array} \right). \end{equation} Letting $v_{\textrm{ew}}^{2} =
v_{u}^{2} + v_{d}^{2} + v_{0}^{2} + v_{\ell}^{2}$ so that
$v_{\textrm{ew}}^{2} = 4M_{Z}^{2}/(g^{2} + g'^{ \hspace{.02 in} 2})
\approx (246 \ \textrm{GeV})^{2}$, we define the mixing angles
$\alpha$, $\beta$, and $\beta_{\ell}$ by the relations $\tan\beta =
v_{u}/v_{d}$, $\tan\beta_{\ell} = v_{0}/v_{\ell}$, and
$\tan^{2}\alpha = (v_{u}^{2} + v_{d}^{2})/(v_{0}^{2} +
v_{\ell}^{2})$. These definitions lead to the following
parameterization of the Higgs vevs: \begin{equation} \begin{split}
\label{eq:vev_breakdowns} v_{u} = v_{\textrm{ew}} \sin\alpha
\sin\beta, \hspace{.2 in} & v_{d} = v_{\textrm{ew}} \sin\alpha
\cos\beta, \\ v_{0} = v_{\textrm{ew}} \cos\alpha \sin\beta_{\ell},
\hspace{.2 in} & v_{\ell} = v_{\textrm{ew}} \cos\alpha
\cos\beta_{\ell}. \end{split} \end{equation}

In order to avoid increasing the $Z$ width or violating other known
bounds, we want the light dark matter to separate from the other
neutralinos and be mostly singlino $\tilde{s}$, the fermionic
component of the singlet $\widehat{S}$. This is accomplished by
taking the parameters $\kappa_{q}$ and $\kappa_{\ell}$ to be small,
which eliminates most of the mixing between the singlino and the
Higgsinos [see Eq. (\ref{eq:neutralinomassmatrix})]. It can then
be easily arranged to have the singlino be the lightest of the
neutralinos. A possible
mechanism for explaining the small size of $\kappa_{q}$ and
$\kappa_{\ell}$ is discussed in Appendix \ref{sec:app2}. Small values of $\kappa_{q}$ and $\kappa_{\ell}$ also
leads to reduced mixing between the scalar singlet and the Higgs
doublets as can be seen from Eq. (\ref{eq:higgspotential}). A small
amount of mixing is of course required since we desire the lightest
scalar, which is mostly singlet, to couple to $\tau$-pairs in order
for the dark matter to annihilate to $\tau^{+}\tau^{-}$ and other
Standard Model particles. This mixing is generated by the soft
supersymmetry-breaking parameters $\mu_{a}$, $\mu_{b}$, $\mu_{c}$,
and $\mu_{d}$.

It is sufficient for $\kappa_{q}$ and $\kappa_{\ell}$ to be
$\mathcal{O}(10^{-2})$, which is what we use in our numerical
calculations (see Table \ref{tab:example} and \ref{tab:example2}). Though the scalar mass
matrices are quite complicated in general, they simplify
considerably in the limit of vanishing $\kappa_{q}$ and
$\kappa_{\ell}$. The numerical calculations in the sections that
follow have been determined using the general matrices, but for
compactness we present only the simplified matrices here. In the
$\{h_{u},h_{d},h_{0},h_{\ell},h_{s}\}$ basis, the neutral scalar
mass matrix is given by \begin{equation} \label{eq:scalarmassmatrix}
M_{N}^{2} = \left( \begin{array}{cc} M^{2} & \overrightarrow{m}^{2} \\
\left.\overrightarrow{m}^{2}\right.^{T} & M_{SS}^{2} \\ \end{array}
\right), \end{equation} where the matrix $M^{2}$ is given by $M^{2}
= M_{\textrm{SLHM}}^{2} + \Delta M_{1}^{2} + \Delta M_{2}^{2}$ and
the terms $\overrightarrow{m}^{2}$ and $M_{SS}$ are given by
\[\left.\overrightarrow{m}^{2} \right.^{T} = -\frac{1}{\sqrt{2}}
\hspace{.03 in} \big(\mu_{a}v_{d} + \mu_{c}v_{\ell}, \hspace{.03 in}
\mu_{a}v_{u} + \mu_{d}v_{0}, \hspace{.03 in} \mu_{b}v_{\ell} +
\mu_{d}v_{d}, \hspace{.03 in} \mu_{b}v_{0} + \mu_{c}v_{u}\big)\] and
\[M_{SS}^{2} = \frac{ 3\big(a_{s} +
\kappa_{s}\lambda_{2}\big)v_{s}^{2} + 2\sqrt{2}
\kappa_{s}^{2}v_{s}^{3} -2t^{3} - 2\lambda_{1}^{2} \lambda_{2} +
\big(\mu_{a}v_{u}v_{d} + \mu_{b}v_{0}v_{\ell} + \mu_{c}v_{u}v_{\ell}
+ \mu_{d}v_{0}v_{d}\big)}{\sqrt{2} \hspace{.03 in} v_{s}}.\] The
matrix $M_{\textrm{SLHM}}^{2}$ is the neutral scalar mass matrix
from the ordinary SLHM, which can be found in
\cite{Marshall:2010qi}, while the matrices $\Delta M_{1}^{2}$ and
$\Delta M_{2}^{2}$ are given by \[ \Delta M_{1}^{2} = \left(
\begin{array}{cccc} -m_{u0}^{2} \frac{v_{0}}{v_{u}} & 0 & m_{u0}^{2}
& 0 \\ 0 & -m_{d\ell}^{2} \frac{v_{\ell}}{v_{d}} & 0 & m_{d\ell}^{2} \\
m_{u0}^{2} & 0 & -m_{u0}^{2} \frac{v_{u}}{v_{0}} & 0 \\ 0 &
m_{d\ell}^{2} & 0 & -m_{d\ell}^{2} \frac{v_{d}}{v_{\ell}}
\\ \end{array} \right), \] and \[ \Delta M_{2}^{2} =  \frac{1}{\sqrt{2}}
\left( \begin{array}{cccc} \frac{v_{s}}{v_{u}} \hspace{.02 in}
(\mu_{a}v_{d} + \mu_{c}v_{\ell}) & -v_{s}\mu_{a} & 0 & -v_{s}\mu_{c} \\
-v_{s}\mu_{a} & \frac{v_{s}}{v_{d}} \hspace{.02 in} (\mu_{a}v_{u} +
\mu_{d}v_{0}) & -v_{s}\mu_{d} & 0 \\ 0 & -v_{s}\mu_{d} &
\frac{v_{s}}{v_{0}} \hspace{.02 in} (\mu_{b}v_{\ell} + \mu_{d}v_{d})
& -v_{s}\mu_{b} \\ -v_{s}\mu_{c} & 0 & -v_{s}\mu_{b} &
\frac{v_{s}}{v_{\ell}} \hspace{.02 in} (\mu_{b}v_{0} + \mu_{c}v_{u}) \\
\end{array} \right).\] The pseudoscalar mass matrix, in the
$\{a_{u},a_{d},a_{0},a_{\ell},a_{s}\}$ basis, is similarly given by
\begin{equation} \label{eq:pseudoscalarmassmatrix} M_{A}^{2} =
\left( \begin{array}{cc} \widetilde{M}^{2} & -\overrightarrow{m}^{2}
\\ -\left. \overrightarrow{m}^{2} \right.^{T} & \widetilde{M}_{SS}^{2} \\
\end{array} \right), \end{equation} where $\widetilde{M}^{2} =
\widetilde{M}_{\textrm{SLHM}}^{2} + \Delta M_{1}^{2} + \Delta
\widetilde{M}_{2}^{2}$. The matrix $\widetilde{M}_{\textrm{SLHM}
}^{2}$ is the pseudoscalar mass matrix from the ordinary SLHM while
$\Delta \widetilde{M}_{2}^{2}$ is the matrix obtained from $\Delta
M_{2}^{2}$ by changing the sign of every off-diagonal entry. Lastly,
$\widetilde{M}_{SS}^{2}$ is given by \begin{align*}
\widetilde{M}_{SS}^{2} = {} & \frac{1}{\sqrt{2} \hspace{.03 in}
v_{s}} \Big[\mu_{a}v_{u}v_{d} + \mu_{b}v_{0}v_{\ell} +
\mu_{c}v_{u}v_{\ell} + \mu_{d}v_{0}v_{d} - 2\lambda_{1}^{2}
\lambda_{2} \\ & - 2t^{3} - \big(9a_{s} +
\kappa_{s}\lambda_{2}\big)v_{s}^{2} - 4\sqrt{2} \hspace{.03 in}
(b_{s}^{2} + \kappa_{s} \lambda_{2}^{2} ) v_{s} \Big].
\end{align*} {The chargino mass matrix, on the other
hand, is rather simple even with nonvanishing $\kappa_{q}$ and
$\kappa_{\ell}$. Letting $\tilde{h}_{u}$, $\tilde{h}_{d}$,
$\tilde{h}_{0}$, and $\tilde{h}_{\ell}$ denote the Higgsino gauge
eigenstates, the chargino mass matrix, in the $\{\widetilde{W}^{+},
\tilde{h}^{+}_{u}, \tilde{h}^{+}_{0}, \widetilde{W}^{-},
\tilde{h}^{-}_{d}, \tilde{h}^{-}_{\ell}\}$ basis, is given by}
\begin{equation} \label{eq:charginomassmatrix} M_{\chi^{\pm}} = \left(
\begin{array}{cccccc}
0 & 0 & 0 & M_{2} & gv_{d} & gv_{\ell} \\
0 & 0 & 0 & gv_{u} & \mu_{q} + \frac{\kappa_{q}}{\sqrt{2}}v_{s} & 0 \\
0 & 0 & 0 & gv_{0} & 0 & \mu_{\ell} + \frac{\kappa_{\ell}}{\sqrt{2}}v_{s} \\
M_{2} & gv_{u} & gv_{0} & 0 & 0 & 0 \\
gv_{d} & \mu_{q} + \frac{\kappa_{q}}{\sqrt{2}}v_{s} & 0 & 0 & 0 & 0 \\
gv_{\ell} & 0 & \mu_{\ell} + \frac{\kappa_{\ell}}{\sqrt{2}}v_{s} & 0 & 0 & 0 \\
\end{array}
\right). \end{equation} {Like the chargino mass
matrix, the neutralino mass matrix is simple. The neutralino mass
matrix, in the $\{\widetilde{B}^{0}, \widetilde{W}^{0},
\tilde{h}_{u}, \tilde{h}_{d}, \tilde{h}_{0}, \tilde{h}_{\ell},
\tilde{s}\}$ basis, is given by} \begin{equation}
\label{eq:neutralinomassmatrix} M_{\chi} = \left(
\begin{array}{ccccccc} M_{1} & 0 & \frac{1}{2} \hspace{.02 in} g'
v_{u} & -\frac{1}{2} \hspace{.02 in} g' v_{d} & \frac{1}{2}
\hspace{.02 in} g' v_{0} & -\frac{1}{2} \hspace{.02 in} g' v_{\ell} & 0 \\
0 & M_{2} & -\frac{1}{2} \hspace{.02 in} g v_{u} & \frac{1}{2}
\hspace{.02 in} g v_{d} & -\frac{1}{2} \hspace{.02 in} g  v_{0} &
\frac{1}{2} \hspace{.02 in} g v_{\ell} & 0 \\
\frac{1}{2} \hspace{.02 in} g' v_{u} & -\frac{1}{2} \hspace{.02 in}
g v_{u} & 0 & \mu_{q} + \frac{\kappa_{q}}{\sqrt{2}} \hspace{.02 in}
v_{s} & 0 & 0 &
\frac{\kappa_{q}}{\sqrt{2}} \hspace{.02 in} v_{d} \\
-\frac{1}{2} \hspace{.02 in} g' v_{d} & \frac{1}{2} \hspace{.02 in}
g v_{d} & \mu_{q} + \frac{\kappa_{q}}{\sqrt{2}} \hspace{.02 in}
v_{s} & 0 & 0 & 0 &
\frac{\kappa_{q}}{\sqrt{2}} \hspace{.02 in} v_{u} \\
\frac{1}{2} \hspace{.02 in} g' v_{0} & -\frac{1}{2} \hspace{.02 in}
g v_{0} & 0 & 0 & 0 & \mu_{\ell} + \frac{\kappa_{\ell}}{\sqrt{2}}
\hspace{.02 in} v_{s} &
\frac{\kappa_{\ell}}{\sqrt{2}} \hspace{.02 in} v_{\ell} \\
-\frac{1}{2} \hspace{.02 in} g' v_{\ell} & \frac{1}{2} \hspace{.02
in} g v_{\ell} & 0 & 0 & \mu_{\ell} + \frac{\kappa_{\ell}}{\sqrt{2}}
\hspace{.02 in} v_{s} & 0 & \frac{\kappa_{\ell}}{\sqrt{2}}
\hspace{.02 in} v_{0} \\ 0 & 0 & \frac{\kappa_{q}}{\sqrt{2}}
\hspace{.02 in} v_{d} & \frac{\kappa_{q}}{\sqrt{2}} \hspace{.02 in}
v_{u} & \frac{\kappa_{\ell}}{\sqrt{2}} \hspace{.02 in} v_{\ell} &
\frac{\kappa_{\ell}}{\sqrt{2}} \hspace{.02 in} v_{0} &
\lambda_{2} + \sqrt{2} \hspace{.02 in} \kappa_{s}v_{s} \\
\end{array} \right). \end{equation} {When $\kappa_{q}$ and
$\kappa_{\ell}$ are small, the singlino part of the above matrix
separates from the wino, bino, and higgsinos, and the singlino mass
can be well approximated by} \begin{equation}
\label{eq:singlinomass} m_{\chi_{1}} \approx \lambda_{2} + \sqrt{2}
\hspace{.02 in}\kappa_{s} v_s. \end{equation} The $\mathcal{O}(10)$
GeV LSP can be arranged with some tuning of the parameters in order
to achieve a cancelation between $\lambda_{2}$ and the product
$\kappa_{s}v_{s}$ in Eq. (\ref{eq:singlinomass}). Though the
smallness of $\kappa_{q}$ and $\kappa_{\ell}$ is technically
unnatural, we remind the reader that a possible mechanism to make
them small is discussed in Appendix \ref{sec:app2}.

In the following sections, we calculate the relevant cross sections
and quantities of interest using benchmark points A and B, found in
Tables \ref{tab:example} and \ref{tab:example2} respectively. While
both of these benchmark points can explain the Galactic Central
region gamma ray excess, the spin independent direct detection cross
section corresponding to benchmark point A lies within the region
favored by CoGeNT and DAMA. In contrast, we will show that benchmark
point B satisfies CDMS bounds that exclude CoGeNT and DAMA. Relevant
quantities have been calculated for several additional benchmark
points as well, and their values are summarized in Table
\ref{tab:benchmarks} of Appendix \ref{sec:app1}.

\begin{table}
\centering
\begin{tabular}{|ccc|ccc|ccc|ccc|ccc|}
\hline \hline $\kappa_{q}$ & $=$ & 0.01 & \ $v_{s}$ & $=$ & 50 GeV &
\ $\mu_{\ell}$ & $=$ & 125 GeV & \ $m_{d\ell}^{2}$ & $=$ & $(100 \
\textrm{GeV})^{2}$ & \ $\mu_{b}$ & $=$ & 200 GeV \\ $\kappa_{\ell}$
& $=$ & 0.01 & \ $v_{u}$ & $=$ & 245.6 GeV & \ $\lambda_{1}^{2}$ &
$=$ & $(100 \ \textrm{GeV})^{2}$ & $\mu_{1}^{2}$ &=& $(400 \ \textrm{GeV})^{2}$ \ & \ $\mu_{c}$ &=& 200 GeV \\
$\kappa_{s}$ & $=$ & 0.6 & \ $v_{d}$ & $=$
& 4.9 GeV & \ $\lambda_{2}$ & $=$ & $-35$ GeV & $\mu_{2}^{2}$ &=& $(200 \ \textrm{GeV})^{2}$ & \ $\mu_{d}$ & $=$ & 200 GeV \\
$\tan\alpha$ & $=$ & 20 & \ $v_{0}$ & $=$ & 12.2 GeV & \ $M_{1}$ &
$=$ & 500 GeV & $\mu_{3}^{2}$ &=& $(200 \ \textrm{GeV})^{2}$ & \ $t^{3}$ & $=$ & $(60.6 \ \textrm{GeV})^{3}$ \ \\
$\tan\beta$ & $=$ & 50 & \ $v_{\ell}$ & $=$ & 1.2 GeV & \ $M_{2}$ & $=$ & 500 GeV & $\mu_{4}^{2}$ &=& $(400 \ \textrm{GeV})^{2}$ & \ $b_{s}^{2}$ & $=$ & $(63.4 \ \textrm{GeV})^{2}$ \ \\
$\tan\beta_l$ & $=$ & 10 & \ $\mu_{q}$ & $=$ & 125 GeV & $ \
m_{u0}^{2}$ & $=$ & $-(100 \ \textrm{GeV})^{2}$ & $\mu_{a}$ & $=$ & 100 GeV & \ $a_{s}$ & $=$ & $-42.4$ GeV \\
\hline \hline
\end{tabular} \caption{Benchmark Point A} \label{tab:example}
\end{table}

\begin{table}
\centering
\begin{tabular}{|ccc|ccc|ccc|ccc|ccc|}
\hline \hline $\kappa_{q}$ & $=$ & 0.01 & \ $v_{s}$ & $=$ & 50 GeV &
\ $\mu_{\ell}$ & $=$ & 125 GeV & \ $m_{d\ell}^{2}$ & $=$ & $(100 \
\textrm{GeV})^{2}$ & \ $\mu_{b}$ & $=$ & 200 GeV \\ $\kappa_{\ell}$
& $=$ & 0.01 & \ $v_{u}$ & $=$ & 245.6 GeV & \ $\lambda_{1}^{2}$ &
$=$ & $(100 \ \textrm{GeV})^{2}$ & $\mu_{1}^{2}$ &=& $(400 \ \textrm{GeV})^{2}$ \ & \ $\mu_{c}$ &=& 200 GeV \\
$\kappa_{s}$ & $=$ & 0.6 & \ $v_{d}$ & $=$
& 4.9 GeV & \ $\lambda_{2}$ & $=$ & $-35$ GeV & $\mu_{2}^{2}$ &=& $(200 \ \textrm{GeV})^{2}$ & \ $\mu_{d}$ & $=$ & 200 GeV \\
$\tan\alpha$ & $=$ & 20 & \ $v_{0}$ & $=$ & 12.2 GeV & \ $M_{1}$ &
$=$ & 500 GeV & $\mu_{3}^{2}$ &=& $(200 \ \textrm{GeV})^{2}$ & \ $t^{3}$ & $=$ & $(55.0 \ \textrm{GeV})^{3}$ \ \\
$\tan\beta$ & $=$ & 50 & \ $v_{\ell}$ & $=$ & 1.2 GeV & \ $M_{2}$ & $=$ & 500 GeV & $\mu_{4}^{2}$ &=& $(400 \ \textrm{GeV})^{2}$ & \ $b_{s}^{2}$ & $=$ & $(66.3 \ \textrm{GeV})^{2}$ \ \\
$\tan\beta_l$ & $=$ & 10 & \ $\mu_{q}$ & $=$ & 125 GeV & $ \
m_{u0}^{2}$ & $=$ & $-(100 \ \textrm{GeV})^{2}$ & $\mu_{a}$ & $=$ & 100 GeV & \ $a_{s}$ & $=$ & $-42.2$ GeV \\
\hline \hline
\end{tabular} \caption{Benchmark Point B} \label{tab:example2}
\end{table}

\section{Annihilation to Fermions}\label{sec:annihilation}
In this section, we will show that this model can achieve the conditions needed to explain the gamma ray excess in the Galactic Center region. In order to calculate the dark matter cross section, we need the interactions between Higgs and fermions:
\begin{equation}\label{eq:singletint} \begin{split}
\lag \supset & -\frac{\kappa_s }{\sqrt{2}} \lf[ h_s\bar{\tilde s} \tilde s - i a_s \bar{\tilde s} \gamma^5\tilde s \rh] \\ &  -\frac{\kappa_q}{2\sqrt{2}} \lf[ h_u\bar{\tilde s} \tilde h_d - i a_u \bar{\tilde s} \gamma^5\tilde h_d +  h_d\bar{\tilde s} \tilde h_u - i a_d \bar{\tilde s} \gamma^5\tilde h_u + h.c. \rh] \\ & -\frac{ \kappa_\ell}{2\sqrt{2}} \lf[ h_0\bar{\tilde s} \tilde h_\ell - i a_0 \bar{\tilde s} \gamma^5\tilde h_\ell +  h_\ell\bar{\tilde s} \tilde h_0 - i a_\ell \bar{\tilde s} \gamma^5\tilde h_0 + h.c. \rh] \\ & - \sum_{f = \lf\{u,d,\ell\rh\}}\sum_j \frac{m_{f_j}}{v_f} \lf(h_f \bar f_jf_j -ia_f \bar f_j\gamma^5 f_j \rh), \end{split}
\end{equation}
where
$m_{f_j}$ is the mass of the fermion $f_j$, $v_f$ is the vev of $f$-type scalars, and $j$ runs over the fermion generations. In the limit $\kappa_q, \kappa_\ell \rightarrow 0,$ the higgs-higgsino-singlino interactions vanish.

\begin{figure}
    \centering
        \includegraphics{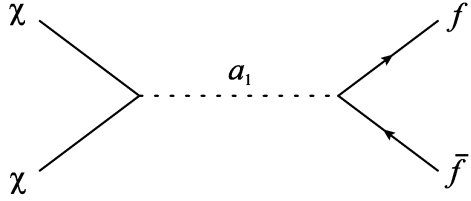}
    \caption{The dominant diagram of dark matter annihilation into fermions. Here $a_1$ is the lightest pseudoscalar.}
    \label{fig:chi}
\end{figure}

We can expand $\langle\sigma v\rangle$ in powers of the dark matter
velocity squared $v^2$: \be \label{eq:expand} \langle\sigma v
\rangle = a + bv^2 + \ldots. \ee {Only the $s$-wave
contribution to $a$ is relevant in discussing the gamma ray excess
coming from dark matter annihilation since the velocity of the dark
matter in the Galactic Center region is relatively low. An exception
to this is within the sphere of influence of the Milky Way
supermassive black hole, but this region corresponds to only a
fraction of an arc second and is below FGST accuracy.} As we see
later, $a_1$ is mostly singlet for benchmark points A and B.
Therefore the $s$-wave contribution to dark matter annihilation to
fermions comes mostly from the $s$-channel diagram involving an
exchange of the lightest pseudoscalar $a_1$ given in Fig.
\ref{fig:chi}. It is approximately given by
\begin{equation}
a \:\approx\: \frac{N_c \kappa_s^2 \: U_{1f}^2 }{4\pi}\frac{m_f^2}{v_f^2} \frac{ m_{\chi_1}^2} {(4m_{\chi_1}^2-m_{a_1}^2)^2}\sqrt{1-\frac{m_f^2}{m_{\chi_1}^2}},
\end{equation}
where $N_c$ is the number of fermion colors, $U_{1f}$ is the $(1,f)$
element of the pseudoscalar diagonalizing matrix and $m_{a_1}$ is
the mass of the lightest pseudoscalar. The $s$-wave contributions
from heavier pseudoscalars  are suppressed by larger masses as well
as smaller mixings with the singlet. Moreover, $s$-channel scalar
exchange diagrams are $s$-wave suppressed, i.e. $a \: (\chi_1\chi_1
\rightarrow h_i \rightarrow \bar ff) = 0$.

For benchmark point A, the dark matter mass is $m_{\chi_1} = 7.4
\textrm{ GeV}$. The physical dark matter can be expressed in terms
of gauge eigenstates as: \be \chi_1 = 0.0017 \: \widetilde B^0 -
0.0031 \: \widetilde W^0 -0.0141 \: \tilde h_u -0.0046 \: \tilde h_d
- 0.0001 \: \tilde h_0 - 0.0008 \: \tilde h_\ell + 0.9999 \: \tilde
s. \nonumber \ee We need a light pseudoscalar, $\mathcal O(10)$ GeV,
to get a sizeable annihilation cross section. This requires 1\%
tuning in the parameter space in addition to the tuning needed to
make the singlino the LSP. 
The lightest pseudoscalar in the benchmark
point is mostly singlet with a mixing with other types of
pseudoscalar given by \be a_1 = - 0.000002 \: a_u - 0.002193 \: a_d
-0.001203 \: a_0 - 0.003679 \: a_\ell + 0.999990 \: a_s \nonumber,
\ee with its mass is $m_{a_1} = 18.7 \textrm{ GeV}$.

Having the masses and mixing, we can calculate the total annihilation cross section into fermion pairs which gives
\begin{equation} \label{eq:dmtot}
\langle\sigma v\rangle \; = \; 4.0 \times 10^{-26} \; \textrm{cm}^3/\textrm{s}
\end{equation}
where the hadronic final states cross section is $23\%$ of the total cross section and $\tau$ pairs final state makes up the rest. For benchmark point B given in Table  \ref{tab:example2}, the mass of dark matter is $m_{\chi_1} = 7.4 $ GeV and $\langle\sigma v\rangle = 3.0 \times 10^{-26} \textrm{ cm}^3/\textrm{s}$, with the hadronic final states make up 23\% of it. The annihillation cross sections given above are within the range of suggested cross section for explaining the gamma ray excess in the Galactic Center region given in Ref. \cite{Hooper:2010mq}.

In this model, dark matter annihilation into SM fermions given in Fig. \ref{fig:chi} is also responsible for giving the dark matter the correct thermal relic abundance. To show this, we calculate the relic abundance which is given by \cite{Jungman:1995df}
\be
\Omega_{\chi_1} h^2 \approx 2.82 \times 10^8 \: Y_\infty (m_{\chi_1}/\textrm{GeV}),
\ee
where
\be
Y_\infty^{-1} = 0.264 \: \sqrt{g_*}m_{P}m_{\chi_1}\left\{a/x_f+3(b-\tfrac{1}{4}a)/x_f^2\right\}.
\ee
In the equation above, $m_{P}$ is the Planck mass and $g_*$ is the number of relativistic degrees of freedom at freeze-out. The freeze-out epoch $x_f$ is related to the freeze-out temperature $T_f$ by $x_f = m_{\chi_1}/T_f$, and $x_f$ is determined by \cite{Jungman:1995df}
\be
x_f = \ln\lf[0.0764\:m_{P}(a+6b/x_f)c(2+c)m_{\chi_1}/\sqrt{g_*x_f}\rh].
\ee
The value of $c$ is usually taken as $c=\tfrac{1}{2}$. Approximating $g_*$ to be a ladder function, we get that, for both of our benchmark points, the freeze-out epoch is $x_f = 21$ and the relic abundance is
\be \label{eq:relic}
\Omega_{\chi_1} h^2 \approx 0.1,
\ee
which agrees with the cosmologically measured abundance \cite{Komatsu:2008hk}. Since the freeze-out temperature happens to be around the QCD phase transition temperature, $g_*$ varies significantly over the change of temperature \cite{Gondolo:1990dk} and the result (\ref{eq:relic}) can change up to $\mathcal O(1)$. However the relic density is in the correct ballpark, therefore we do not expect that the correction will invalidate our result. An adjustment of parameters can be done when taking into account of the variation of $g_*$ to get the correct density and annihilation cross section.

The benchmark points A and B serve as examples to show that in
principle this model can explain the gamma ray excess in the
Galactic Center region. However, the excess could also be obtained
by some other regions in the parameter space as shown in the
Appendix \ref{sec:app1}. One could do a scan on the parameter space
to find the favored region of the model.

{Note that
in our relic density calculation, we have neglected possible
chargino and sfermion contributions coming from resonance and coannihilation effects. This is because the charginos have masses
$\mathcal{O}(100)$ GeV for all of our benchmark points, and we
assume that the sfermion masses are at least $\mathcal{O}(100)$ GeV,
which is consistent with current LEP bounds.}

\section{Direct Detection}\label{sec:direct}
Having shown that this model can account for the gamma ray excess in the Galactic Center region, we now discuss direct detection of dark matter of this model. In this section, we will consider constraints from the search for spin independent, elastic scattering of dark matter off target nuclei. The most relevant contribution for the cross section is given by the $t$-channel scalar exchange diagram with the effective Lagrangian:
\be
\mathcal L_{int} = \sum_q \alpha_q \bar\chi_1\chi_1\bar qq.
\ee
In our benchmark points, the only relevant contribution to dark matter detection comes from the lightest scalar and $\alpha_q$ can be approximated by
\be
\alpha_q \approx \frac{\kappa_s m_q V_{1q}}{\sqrt{2} v_q m_{h_1}^2},
\ee
where $m_q$ is the mass of quark $q$, $v_q$ is the scalar vev associated with quark flavor $q$, $V_{1q}$ is the $(1,q)$ element of the scalar diagonalizing matrix, and $m_{h_1}$ is the mass of the lightest scalar. Given the partonic interaction between dark matter and quarks, we can follow Ref.~\cite{Munoz:2003gx} to get the effective interaction with nucleons:
\begin{equation}
\mathcal L_{eff} = f_p  \,\bar\chi_1\chi_1 \, \bar pp + f_n  \,\bar\chi_1\chi_1 \, \bar nn ,
\end{equation}
where $f_p$ and $f_n$ are related to $\alpha_q$ through the relation~\cite{Munoz:2003gx}
\begin{equation}
\frac{f_{p,n}}{m_{p,n}} = \displaystyle\sum_{q=u,d,s} \frac{f^{(p,n)}_{Tq}\alpha_q}{m_q}
+ \frac{2}{27} \, f_{Tg}^{(p,n)}\displaystyle\sum_{q=c,b,t}\frac{\alpha_q}{m_q},
\end{equation}
and $\langle n|m_q\bar qq|n \rangle =m_nf_{Tq}^n$. Numerically, the $f^{(p,n)}_{Tq}$
are given by~\cite{Ellis:2000ds}
\be \begin{split}
f_{Tu}^p=0.020\pm0.004,\; & f_{Td}^p=0.026\pm0.005,\; f_{Ts}^p = 0.118\pm0.062 \\
f_{Tu}^n=0.014\pm0.0043,\; & f_{Td}^n=0.036\pm0.008,\; f_{Ts}^n = 0.118\pm0.062 ,
\end{split}
\ee
while $f_{Tg}^{(p,n)}$ is defined by
\begin{equation}
f_{Tg}^{(p,n)}=1-\displaystyle\sum_{q=u,d,s}f_{Tq}^{(p,n)}\,\, .
\end{equation}
We can approximate $f_p \approx f_n$ since $f_{Ts}$ is larger than other $f_{Tq}$'s and $f_{Tg}$. For
the purpose of comparing the predicted cross section with existing bounds, we evaluate the
cross section for scattering off a single nucleon. The result can be approximated as
\begin{equation}
\sigma_{SI} \approx \frac{4 m_r^2 f_p^2}{\pi}
\end{equation}
where $m_r$ is nucleon-dark matter reduced mass $1/m_r = 1/m_n+1/m_{\chi_1}$.

We are now ready to show that benchmark point A can explain signals reported by CoGeNT \cite{Aalseth:2010vx} and DAMA \cite{Bernabei:2008yi}. 
For this benchmark point, the lightest scalar mass is $m_{h_1} = 11.3 \textrm{ GeV}$. This lightest scalar is mostly singlet and its mixing with other scalars is given by
\be
h_1 = 0.089 \: h_u + 0.004 \: h_d + 0.010 \: h_0 + 0.004 \: h_\ell + 0.996 \: h_s. \nonumber
\ee
As in the case of pseudoscalar, contributions from higher mass scalars are suppressed by their masses and their mixings with the singlet. The spin independent cross section for the benchmark point now can be calculated and is given by
\be
\sigma_{SI} = 1.7 \times 10^{-40} \textrm{ cm}^2,
\ee
which is inside the CoGeNT and DAMA favored region \cite{Hooper:2010uy}.

Similarly, we can show that benchmark point B given in Table \ref{tab:example2} has the lightest scalar mass $m_{h_1} = 41.5$ GeV and spin independent cross section $\sigma_{SI} = 1.2 \times 10^{-42} \textrm{ cm}^2$. This cross section is two orders of magnitude lower than the present CDMS and XENON bound \cite{Aprile:2010um,Ahmed:2010wy}.


\section{Bounds on the Model} \label{sec:bounds}

In this section we discuss various collider bounds that apply to the model. We will spend most of the discussions in this section for the benchmark point A given in Table \ref{tab:example}. The bounds for benchmark point B as well as the summary of the bounds for benchmark point A are given in Table \ref{table:bounds}.

In this model, the decays $Z \rightarrow \chi_1\chi_1$ and $Z
\rightarrow h_1a_1$ are allowed kinematically. The $Z$ decay width
has been measured precisely and is given by $\Gamma = 2.4952 \pm
0.0023$ GeV \cite{Nakamura:2010zzi}. Corrections to the decay width
can be used as a bound on the mixing between the singlet and the
Higgs sector. The partial decay width of $Z\rightarrow \chi_1\chi_1$
is given by \be \Gamma_{Z\rightarrow \chi_1\chi_1} =
\frac{G_F\theta_\chi^2}{48\sqrt{2}\pi}m_Z^3\lf(1-\frac{4m_{\chi_1}^2}{m_Z^2}\rh)^{\tfrac{3}{2}},
\ee where $G_F$ is the Fermi constant, $m_Z$ is $Z$ mass, and
$\theta_\chi$ is given by \be \theta_\chi = \lf|W_{u1}\rh|^2 -
\lf|W_{d1}\rh|^2  + \lf|W_{01}\rh|^2  - \lf|W_{\ell 1}\rh|^2 . \ee
In the equation above, $W_{f1}$ is the $(f,1)$ element of the
neutralino diagonalizing matrix. The decay width of $Z \rightarrow
h_1a_1$ is given by \be \Gamma_{Z\rightarrow h_1a_1} =
\frac{G_F|\theta_{ha}|^2}{3\sqrt{2}\pi}p^3, \ee where \be
\label{eq:UV} \theta_{ha} = U_{u1}V_{u1} - U_{d1}V_{d1} +
U_{01}V_{01} - U_{\ell 1}V_{\ell 1}, \ee and \be p^2 =
\frac{1}{4m_Z^2}\lf[\lf(m_Z^2-\lf(m_{h_1}+m_{a_1}\rh)^2\rh)\lf(m_Z^2-\lf(m_{h_1}-m_{a_1}\rh)^2\rh)\rh].
\ee For the benchmark point, the partial decay widths in both cases
are given by \be \begin{split}
\Gamma_{Z\rightarrow \chi_1\chi_1} & = 1.4 \times 10^{-9} \textrm{ GeV}, \\
\Gamma_{Z\rightarrow h_1a_1} & = 1.1 \times 10^{-11} \textrm{ GeV}, \end{split}
\ee
which is well within the measurement error.

\begin{table}
\centering
\begin{tabular}{|c|c|c|c|c|c|c|}
\hline \hline
Benchmark point & A & B \\
\hline
$m_{\chi_1}$ (GeV) & 7.4 & 7.4 \\
$m_{\chi_{1}^{\pm}}$ (GeV) & 118 & 118 \\
$m_{h_1}$ (GeV) & 11.3 & 41.5 \\
$m_{a_1}$ (GeV) & 18.7 & 19.3 \\
$\Gamma_{Z \rightarrow \chi_1\chi_1}$ (GeV) & $1.4 \times 10^{-9}$ & $1.4 \times 10^{-9}$ \\
$\Gamma_{Z \rightarrow h_1 a_1}$ (GeV) & $1.1 \times 10^{-11}$ & $4.9 \times 10^{-12}$ \\
$k$ & $8.0 \times 10^{-3}$ &  $1.3 \times 10^{-2}$ \\
$S_{model} (e^+e^- \rightarrow h_1a_1)$ & $1 \times 10^{-10}$ & $1 \times 10^{-10}$ \\
$S_{model} (e^+e^- \rightarrow h_2a_1)$ & $1 \times 10^{-12}$ & $2 \times 10^{-12}$ \\
$\sigma_{e^+e^- \rightarrow \chi_1\chi_2}$ (pb) & $1 \times 10^{-5}$ & $1 \times 10^{-5}$ \\
\hline
\hline
\end{tabular} \caption{Mass spectrum and bounds for benchmark points A and B. The variable $k$ is given by $k = \sigma_{hZ}/\sigma_{hZ}^{SM}$ and $S_{model} = \sigma_{h_ia_j}/\sigma_{ref}$, where $\sigma_{h_ia_j}$ is the $h_ia_j$ production cross section and $\sigma_{ref}$ is the reference cross section defined in Ref. \cite{Schael:2006cr}.} \label{table:bounds}
\end{table}

Another bound on the model comes from scalar and pseudoscalar direct production at LEP. At LEP a light scalar can be produced by Higgsstrahlung process $e^+e^- \rightarrow Z \rightarrow Zh_1$. Ref. \cite{Abbiendi:2002qp} gives a bound on the coupling strength of $Z$ pairs to scalars regardless of the scalar's decay mode. The bound is given in terms of the quantity
\be
k(m_h) = \frac{\sigma_{hZ}}{\sigma_{hZ}^{SM}}.
\ee
In our model, $k(m_h)$ is given by
\be
k(m_{h_i}) = \frac{1}{v_{\textrm{ew}}^2} \left| v_u V_{ui} + v_d V_{di} + v_0 V_{0i} + v_\ell V_{\ell i} \right|^2,
\ee
and its value for the lightest scalar at our benchmark point is
\be
k(m_{h_1}) = 8.0\times 10^{-3}.
\ee
The bound on $k({m_h})$ for the benchmark point $h_1$ mass is given by
\be
k(11.3 \textrm{ GeV}) \leq 0.09.
\ee
Therefore $k(m_{h_1})$  does not exceed the bound from Higgsstrahlung process in our benchmark point.
The pseudoscalar can also be produced at LEP by the process $e^+e^- \rightarrow Z \rightarrow ha$. In the benchmark point, both $h_1a_1$ and $h_2a_1$ production are kinematically allowed. LEP bounds on scalar and pseudoscalar production for various final states are given in Ref. \cite{Schael:2006cr}. The bound is given in term of $S_{95} = \sigma_{max}/\sigma_{ref}$ where $\sigma_{max}$ is the largest cross section compatible with data and $\sigma_{ref}$ is the standard model $hZ$ production cross section multiplied by a kinematic scaling factor. Defining $S_{model} = \sigma_{h_ia_j}/\sigma_{ref}$, where $\sigma_{h_ia_j}$ is the model's $h_ia_j$ production cross section, the bound on the model is given by $S_{model} < S_{95}$. For our benchmark point, $S_{model}$ is given by
\be \begin{split}
S_{model}(e^+e^- \rightarrow h_1a_1) &= 1 \times 10^{-10}, \\
S_{model}(e^+e^- \rightarrow h_2a_1) &= 1 \times 10^{-12},
\end{split}
\ee
which is lower than the bound, $S_{95} \sim \mathcal O(10^{-2})$, in both cases.

We note that the lightest chargino mass is 118 GeV for the benchmark
point, which exceeds the PDG bound of 94 GeV
\cite{Nakamura:2010zzi}. In the case of a long lived chargino
however, the bound can be made much stronger and is currently at 171
GeV. We have calculated the lifetime of the chargino in our model
assuming a stau mass of 110 GeV and have found that it is short
lived, thus this latter bound is not of concern. We should point out
however, that our analysis has been done at tree level. Loop
corrections could change these results but are beyond the scope of
this paper.

Finally, we need to calculate the bound on neutralino productions.
Ref. \cite{Abdallah:2003xe} discusses the bound on production of the
lightest and second to lightest neutralinos at LEP, $e^+e^-
\rightarrow \chi_1\chi_2$, where $\chi_2$ decays into $\chi_1 f\bar
f$. Assuming that the selectron is much heavier than the $Z$, the
main contribution comes from s-channel $Z$ exchange. For our
benchmark point, we calculate the cross section to be \be
\sigma_{e^+e^- \rightarrow \chi_1\chi_2} = 1 \times 10^{-5}\textrm{
pb}, \ee while the bound is $\mathcal O(0.1)$ pb. A summary of all
these bounds is given in Table \ref{table:bounds}.

The light particles are mostly singlet and have very little mixing with the Higgs sector. This make the particles unlikely to be produced at near future experiments. However the heavier sector has a richer phenomenology. For example, heavier scalars are mostly $h_u$, $h_d$, $h_0$, and $h_\ell$ therefore they have a better chance of being detected in future colliders \cite{Marshall:2010qi}.

\section{Conclusions} \label{sec:conc}

In this paper, we have presented a supersymmetric model of $7-10$
GeV dark matter, which is capable of describing the FGST
observations. In a recent analysis of FGST data, Hooper and
Goodenough found an excess in gamma ray emission from within
$1.25^{\circ}$ of the Galactic Center. They showed that this can be
explained by annihilating dark matter if the dark matter has a mass
between $7-10$ GeV, annihilates into $\tau$-pairs most of the time,
but into hadronic channels the other $15-40 \%$ of the time, and
$\langle\sigma v\rangle$ falls within the range $4.6 \times 10^{-27} - 5.3
\times 10^{-26} \ \textrm{cm}^{3}/ \textrm{s}$ \cite{Hooper:2010mq}.
Our model achieves these requirements by minimally extending the
SLHM to include a scalar singlet whose superpartner is the dark
matter particle. Due to the Yukawa structure of the SLHM the scalar
particles mediating the dark matter annihilation have an enhanced
coupling to leptons. This provides a natural means for satisfying
the second requirement put forward by Hooper and Goodenough.

We have shown that this model produces the correct dark matter
thermal relic density and is consistent with current collider
bounds. In addition, we have shown that this model is consistent
with the direct detection signals reported by both CoGeNT and DAMA
for certain regions of parameter space, while for other regions of
parameter space, the model yields a spin independent cross section
far below the present CDMS bound, but maintains the right relic
density and continues to explain the FGST observations. Thus our
model is fully able to accommodate the results reported by CoGeNT
and DAMA in the case of their vindication, but it is in no way
contingent upon their validity.

\begin{acknowledgments}
We thank Chris Carone and Marc Sher for useful discussions and their many comments on this manuscript. We also thank Dylan Albrecht for comments on this manuscript. This work was supported by the NSF under Grant PHY-0757481.
\end{acknowledgments}

\appendix

\section{Breaking Terms} \label{sec:app2}

\begin{table}
\centering
\begin{tabular}{|c|c|c|c|c|c|}
\hline \hline \ Field \ & \ $\mathbb{Z}_{3q}$ \ & \
$\mathbb{Z}_{3\ell}$ \ & \ Field \ & \ $\mathbb{Z}_{3q}$ \ & \
$\mathbb{Z}_{3\ell}$ \ \\ \hline
$\widehat{H}_{u}$ & $\omega$ & 1 & $\widehat{X}_{01}$ & 1 & 1 \\
$\widehat{H}_{d}$ & $\omega$ & 1 & $\widehat{X}_{02}$ & $\omega^{2}$ & $\omega^{2}$ \\
$\widehat{H}_{0}$ & 1 & $\omega$ & $\widehat{X}_{q1}$ & $\omega$ & 1 \\
$\widehat{H}_{\ell}$ & 1 & $\omega$ & $\widehat{X}_{q2}$ & $\omega^{2}$ & 1 \\
$\widehat{E}$ & 1 & $\omega^{2}$ & $\widehat{X}_{\ell 1}$ & 1 & $\omega$ \\
$\widehat{Q}$ & $\omega^{2}$ & 1 & $\widehat{X}_{\ell 2}$ & 1 & $\omega^{2}$ \\
\hline \hline \end{tabular} \caption{Transformation rule for the
$\mathbb{Z}_{3q} \times \mathbb{Z}_{3\ell}$ symmetry. Each field
transforms as $\phi \rightarrow X \phi$, where $X$ is the
corresponding factor shown in the table. For each case, $\omega^{3}
= 1$. Other fields not shown in the table are neutral under
$\mathbb{Z}_{3q} \times \mathbb{Z}_{3\ell}$} \label{tab:Z3}
\end{table}

In this appendix, we discuss a possible source of the terms in
$V_{\textrm{soft}}$ that break the $\mathbb{Z}_{2}$ symmetry of the
superpotential. Generally, one can imagine such breaking terms
arising from the $F$-term of some hidden sector superfield receiving
a vacuum expectation value. To be more specific, we consider a
possible scenario that results in such breaking terms and also
explains the smallness of $\kappa_{q}$ and $\kappa_{\ell}$. In this
scenario there is a hidden sector, which contains the six fields
$\widehat{X}_{01}$, $\widehat{X}_{02}$, $\widehat{X}_{q1}$,
$\widehat{X}_{q2}$, $\widehat{X}_{\ell 1}$ and $\widehat{X}_{\ell
2}$. The $F$-terms of the fields receive vevs \begin{equation}
\langle F_{X_{i}} \rangle \sim \mathcal{O}(10^{11} \textrm{GeV})^2,
\end{equation} so that \begin{equation} M_{\textrm{SUSY}} \sim
\frac{\langle F_{X_{i}}\rangle}{M_{P}} \end{equation} is at the TeV
scale. The index $i$ denotes $01$, $02$, $q1$, $q2$, $\ell 1$, and
$\ell 2$. A $\mathbb{Z}_{3q} \times \mathbb{Z}_{3\ell}$ symmetry is
imposed, under which the fields transform according to Table
\ref{tab:Z3}. The hidden sector fields $\widehat{X}_{i}$ couple to
visible sector fields in a high energy, fundamental theory, and are
Planck suppressed in the low energy effective theory.
Consequentially, the lagrangian contains terms such as
\begin{equation} \label{eq:app2eq1} \Delta \mathcal{L} =
\frac{f'}{M_{P}^{2}} \hspace{.02 in} \int d^{4}\theta
\widehat{X}_{01}^{\hspace{.02 in} \dag} \widehat{X}_{02}
\widehat{H}_{u} \widehat{H}_{\ell} + \frac{m'}{M_{P}} \hspace{.02
in} \int d^{2}\theta \widehat{X}_{02} \widehat{S} \widehat{H}_{u}
\widehat{H}_{\ell} + \textrm{h.c.}, \end{equation} where
$d^{2}\theta = d (\theta \theta)$ and $d^{4}\theta = d (\theta
\theta) d (\bar{\theta} \bar{\theta})$ represent integration over
Grassmann variables and $f'$ and $m'$ are coupling constants. When
the $F$-terms of $\widehat{X}_{01}$ and $\widehat{X}_{02}$ receive
vevs, the terms in Eq. (\ref{eq:app2eq1}) give rise to
\begin{equation} \begin{split} \label{eq:app2eq2} \Delta \mathcal{L} = {} &
\frac{f' \langle F_{01} \rangle \langle F_{02} \rangle}{M_{P}^{2}}
\hspace{.02 in} \int d^{4}\theta (\bar{\theta} \bar{\theta}) (\theta
\theta) \widehat{H}_{u} \widehat{H}_{\ell} + \frac{m' \langle F_{02}
\rangle}{M_{P}} \hspace{.02 in} \int d^{2}\theta (\theta \theta)
\widehat{S} \widehat{H}_{u} \widehat{H}_{\ell} + \textrm{h.c.} \\ =
{} & \frac{f' \langle F_{01} \rangle \langle F_{02}
\rangle}{M_{P}^{2}} \hspace{.02 in} H_{u}H_{\ell} + \frac{m' \langle
F_{02} \rangle}{M_{P}}
\hspace{.02 in} \hspace{.02 in} S H_{u} H_{\ell} + \textrm{h.c.} \\
\rightarrow {} & \mu_{3}^{2}H_{u}H_{\ell} + \mu_{c}SH_{u}H_{\ell} +
\textrm{h.c.}. \end{split} \end{equation}

\begin{table}
\centering
\begin{tabular}{|c|c|}
\hline \hline
$\f{a'}{M_P}\int d^{4}\theta \widehat{X}_{q2}^{\;\dagger} \widehat H_u \widehat H_d + \textrm{h.c.}$ & $\int d^{2}\theta \mu_q \widehat H_u \widehat H_d + \textrm{h.c.}$\\
$\f{b'}{M_P}\int d^{4}\theta \widehat{X}_{\ell 2}^{\;\dagger} \widehat H_{0} \widehat H_l + \textrm{h.c.}$ & $\int d^{2}\theta \mu_\ell \widehat H_0 \widehat H_\ell + \textrm{h.c.}$\\
$\f{c'}{M_P}\int d^{4}\theta \widehat{X}_{01}^{\;\dagger} \widehat S^2 + \textrm{h.c.}$ & $\int d^{2}\theta \lambda_2 \widehat S^2 + \textrm{h.c.}$\\
$\f{1}{M_P^2}\int d^{4}\theta \lf(d'\widehat{X}_{01}^{\;\dagger}\widehat{X}_{q1}+d''\widehat{X}_{q2}^{\;\dagger}\widehat{X}_{01}+d'''\widehat{X}_{02}^{\;\dagger}\widehat{X}_{\ell 2}+ d''''\widehat{X}_{q1}^{\;\dagger}\widehat{X}_{q2}\rh) \widehat H_u \widehat H_d + \textrm{h.c.}$ & $ \mu_1^2 H_u H_d + \textrm{h.c.}$\\
$\f{1}{M_P^2}\int d^{4}\theta \lf(e'\widehat{X}_{01}^{\;\dagger}\widehat{X}_{\ell 1}+e''\widehat{X}_{\ell 2}^{\;\dagger}\widehat{X}_{01}+e'''\widehat{X}_{02}^{\;\dagger}\widehat{X}_{q 2}+ e''''\widehat{X}_{\ell 1}^{\;\dagger}\widehat{X}_{\ell 2}\rh) \widehat H_0 \widehat H_\ell + \textrm{h.c.}$ & $ \mu_2^2 H_0 H_\ell + \textrm{h.c.}$\\
$\f{1}{M_P^2}\int d^{4}\theta \lf(f'\widehat{X}_{01}^{\;\dagger}\widehat{X}_{02}+f''\widehat{X}_{q1}^{\;\dagger}\widehat{X}_{\ell 2}+f'''\widehat{X}_{\ell 1}^{\;\dagger}\widehat{X}_{q 2}\rh) \widehat H_u \widehat H_\ell + \textrm{h.c.}$ & $ \mu_3^2 H_u H_\ell + \textrm{h.c.}$\\
$\f{1}{M_P^2}\int d^{4}\theta \lf(g'\widehat{X}_{01}^{\;\dagger}\widehat{X}_{02}+g''\widehat{X}_{q1}^{\;\dagger}\widehat{X}_{\ell 2}+g'''\widehat{X}_{\ell 1}^{\;\dagger}\widehat{X}_{q 2}\rh) \widehat H_0 \widehat H_d + \textrm{h.c.}$ & $ \mu_4^2 H_0 H_d + \textrm{h.c.}$\\
$\f{1}{M_P^2}\int d^{4}\theta \lf(h'\widehat{X}_{02}^{\;\dagger}\widehat{X}_{\ell 1}+h''\widehat{X}_{q1}^{\;\dagger}\widehat{X}_{02}+h'''\widehat{X}_{q2}^{\;\dagger}\widehat{X}_{\ell 2}+ h''''\widehat{X}_{\ell 1}^{\;\dagger}\widehat{X}_{q1}\rh) \widehat H_u^{\; \dagger} \widehat H_0 + \textrm{h.c.}$ & $ m_{u0}^2 H_u^{\dagger} H_0 + \textrm{h.c.}$\\
$\f{1}{M_P^2}\int d^{4}\theta \lf(i'\widehat{X}_{02}^{\;\dagger}\widehat{X}_{\ell 1}+i''\widehat{X}_{q1}^{\;\dagger}\widehat{X}_{02}+i'''\widehat{X}_{q2}^{\;\dagger}\widehat{X}_{\ell 2}+ i''''\widehat{X}_{\ell 1}^{\;\dagger}\widehat{X}_{q1}\rh) \widehat H_d^{\; \dagger} \widehat H_\ell + \textrm{h.c.}$ & $ m_{d\ell}^2 H_d^{\dagger} H_\ell + \textrm{h.c.}$\\
$\f{1}{M_P^2}\int d^{4}\theta \sum_{i} j^{i}\widehat{X}_{i}^{\;\dagger}\widehat{X}_{i} \widehat H_f^{\; \dagger} \widehat H_f + \textrm{h.c.}$ & $ m_f^2  |H_f|^2 + \textrm{h.c.}$\\
$\f{k'}{M_P}\int d^{2}\theta \widehat{X}_{q1} \widehat S \widehat H_u \widehat H_d + \textrm{h.c.}$ & $ \mu_a S H_u H_d + \textrm{h.c.}$\\
$\f{l'}{M_P}\int d^{2}\theta \widehat{X}_{\ell 1} \widehat S \widehat H_0 \widehat H_\ell + \textrm{h.c.}$ & $ \mu_b S H_0 H_\ell + \textrm{h.c.}$\\
$\f{m'}{M_P}\int d^{2}\theta \widehat{X}_{02} \widehat S \widehat H_u \widehat H_\ell + \textrm{h.c.}$ & $ \mu_c S H_u H_\ell + \textrm{h.c.}$\\
$\f{n'}{M_P}\int d^{2}\theta \widehat{X}_{02} \widehat S \widehat H_0 \widehat H_d + \textrm{h.c.}$ & $ \mu_d S H_0 H_d + \textrm{h.c.}$\\
$\f{1}{M_P^2}\int d^{4}\theta \sum_{i} o^{i}\widehat{X}_{i}^{\;\dagger}\widehat{X}_{i} \widehat S^2 + \textrm{h.c.}$ & $ b_s^2 S^2 + \textrm{h.c.}$\\
$\f{p'}{M_P}\int d^{2}\theta \widehat{X}_{0} \widehat S^3 + \textrm{h.c.}$ & $ a_s S^3 + \textrm{h.c.}$\\
\hline \hline
\end{tabular} \caption{A complete list of superpotential and $V_{\textrm{soft}}$ terms generated by the $X_{i}$ in this example.} \label{tab:breaking}
\end{table}

Similarly, the breaking
parameters $\mu_{4}^{2}$ and $\mu_{d}$ arise from the Planck
suppressed terms \begin{equation} \begin{split} \label{eq:app2eq3}
\Delta \mathcal{L} = {} & \frac{g'}{M_{P}^{2}} \hspace{.02 in} \int
d^{4}\theta \widehat{X}_{01}^{\hspace{.02 in} \dag} \widehat{X}_{02}
\widehat{H}_{0} \widehat{H}_{d} + \frac{n'}{M_{P}} \hspace{.02 in}
\int d^{2}\theta \widehat{X}_{02} \widehat{S} \widehat{H}_{0}
\widehat{H}_{d} + \textrm{h.c.} \\ \rightarrow {} & \frac{g' \langle
F_{01} \rangle \langle F_{02} \rangle}{M_{P}^{2}} \hspace{.02 in}
H_{0}H_{d} + \frac{n' \langle F_{02} \rangle}{M_{P}} \hspace{.02 in}
\hspace{.02 in} S H_{0} H_{d} + \textrm{h.c.} \\ \rightarrow {} &
\mu_{4}^{2}H_{0}H_{d} + \mu_{d}SH_{0}H_{d} + \textrm{h.c.},
\end{split} \end{equation} while the parameters $m_{u0}^{2}$ and $m_{d\ell}^{2}$
arise from \begin{equation} \begin{split} \label{eq:app2eq4} \Delta
\mathcal{L} = {} & \frac{h'}{M_{P}^{2}} \hspace{.02 in} \int
d^{4}\theta \widehat{X}_{02}^{\hspace{.02 in} \dag}
\widehat{X}_{\ell 1} \widehat{H}_{u}^{\hspace{.02 in} \dag}
\widehat{H}_{0} + \frac{i'}{M_{P}^{2}} \hspace{.02 in} \int
d^{4}\theta \widehat{X}_{02}^{\hspace{.02 in} \dag}
\widehat{X}_{\ell 1} \widehat{H}_{d}^{\hspace{.02 in} \dag}
\widehat{H}_{\ell} + \textrm{h.c.}\\ \rightarrow {} & \frac{h'
\langle F_{02} \rangle \langle F_{\ell 1} \rangle}{M_{P}^{2}}
\hspace{.02 in} H_{u}^{\dag}H_{0} + \frac{i' \langle F_{02} \rangle
\langle F_{\ell 1} \rangle}{M_{P}^{2}} \hspace{.02 in}
H_{d}^{\dag}H_{\ell} + \textrm{h.c.} \\ \rightarrow {} & m_{u0}^{2}
H_{u}^{\dag} H_{0} + m_{d\ell}^{2} H_{d}^{\dag} H_{\ell} +
\textrm{h.c.}. \end{split}
\end{equation}

In this way, all of the $\mathbb{Z}_{2}$ breaking terms are
generated. At this point it should be noted that the $\mathbb{Z}_{3q}
\times \mathbb{Z}_{3\ell}$ symmetry actually prohibits the terms
$\mu_{q} \hat{H}_{u} \hat{H}_{d}$, $\mu_{\ell} \hat{H}_{0}
\hat{H}_{\ell}$, $\kappa_{q} \hat{S} \hat{H}_{u} \hat{H}_{d}$, and
$\kappa_{\ell} \hat{S} \hat{H}_{0} \hat{H}_{\ell}$ from appearing in
the superpotential [see Eq. (\ref{eq:superpotential})]. As far as
the $\mu_{q}$ and $\mu_{\ell}$ terms are concerned, this is not a
problem since they are generated by the vevs of the
$\widehat{X}_{q2}$ and $\widehat{X}_{\ell 2}$ fields in the same
manner: \begin{equation} \begin{split} \label{eq:app2eq5} \Delta
\mathcal{L} = {} & \frac{a'}{M_{P}} \hspace{.02 in} \int d^{4}\theta
\widehat{X}_{q2}^{\hspace{.02 in \dag}} \widehat{H}_{u}
\widehat{H}_{d} + \frac{b'}{M_{P}} \hspace{.02 in} \int d^{4}\theta
\widehat{X}_{\ell 2}^{\hspace{.02 in \dag}} \widehat{H}_{0}
\widehat{H}_{\ell} \\ \rightarrow {} & \frac{a' \langle F_{q2}
\rangle}{M_{P}} \int d^{2}\theta d^{2}\bar{\theta} \hspace{.02 in}
(\bar{\theta} \bar{\theta}) \widehat{H}_{u} \widehat{H}_{d} +
\frac{b' \langle F_{\ell 2} \rangle}{M_{P}} \int d^{2}\theta
d^{2}\bar{\theta} \hspace{.02 in} (\bar{\theta} \bar{\theta})
\widehat{H}_{0} \widehat{H}_{\ell} \\ = {} & \frac{a' \langle F_{q2}
\rangle}{M_{P}} \int d^{2}\theta \widehat{H}_{u} \widehat{H}_{d} +
\frac{b' \langle F_{\ell 2} \rangle}{M_{P}} \int d^{2}\theta
\widehat{H}_{0} \widehat{H}_{\ell} \\ \rightarrow {} & \mu_{q} \int
d^{2}\theta \widehat{H}_{u} \widehat{H}_{d} + \mu_{\ell} \int
d^{2}\theta \widehat{H}_{0} \widehat{H}_{\ell}.
\end{split} \end{equation} In this UV completion scenario, the terms
corresponding to $\kappa_q$, $\kappa_\ell$, $\lambda_1$ and $t$ are
not generated in this way. Because of the $\mathbb{Z}_{3q} \times
\mathbb{Z}_{3\ell}$ symmetry, they are entirely absent at tree
level. Benchmark points II and V in Table \ref{tab:benchmarks}
satisfy $\kappa_q = \kappa_\ell = \lambda_1 = t = 0$ and yield
results consistent with our goals. Since we are not committing to
this particular UV completion scheme, we consider several other
benchmark points that include nonzero values for these parameters. A
list of the soft breaking terms relevant to this paper, which are
generated by the fields $X_{i}$, is given in Table
\ref{tab:breaking}.

\section{List of benchmark points} \label{sec:app1}
In this Appendix, we show several benchmark points given in Table \ref{tab:benchmarks}. Benchmarks point I-III lie in the suggested CoGeNT and DAMA range, while benchmarks point IV-VI satisfy CDMS bound. Benchmark point I is identical with benchmark point A discussed in the text. Benchmark point IV is identical with benchmark point B. Benchmark points II and V are motivated by mechanism described in Appendix \ref{sec:app2}.

\begin{table}
\centering \caption{Additional benchmark points}  \label{tab:benchmarks}
\begin{tabular}{|c|c|c|c|c|c|c|}
\hline \hline
Benchmark point & I & II & III & IV & V & VI\\
\hline
$\kappa_{q}$ & 0.01 & 0 & 0.01 & 0.01 & 0 & 0.01 \\
$\kappa_{l}$ & 0.01 & 0 & 0.01 & 0.01 & 0 & 0.01 \\
$\kappa_{s}$ & 0.6 & 0.6 & 0.5 & 0.6 & 0.6 & 0.5 \\
$\tan\alpha$ & 20 & 15 & 30 & 20 & 30 & 25 \\
$\tan\beta$ & 50 & 30 & 30 & 50 & 25 & 25 \\
$\tan\beta_\ell$ & 10 & 10 & 5 & 10 & 5 & 5 \\
$v_s$ (GeV) & 50 & 50 & 100 & 50 & 50 & 100 \\
$v_u$ (GeV) & 245.6 & 245.3 & 245.7 & 245.6 & 245.7 & 245.6 \\
$v_d$ (GeV) & 4.9 & 8.2 & 8.2 & 4.9 & 9.8 & 9.8 \\
$v_0$ (GeV) & 12.2 & 16.2 & 8.0 & 12.2 & 8.0 & 9.6 \\
$v_\ell$ (GeV) & 1.2 & 1.6 & 1.6 & 1.2 & 1.6 & 1.9 \\
$\mu_q$ (GeV) & 125 & 125 & 200 & 125 & 125 & 150 \\
$\mu_\ell$ (GeV) & 125 & 125 & 150 & 125 & 150 & 150 \\
$\lambda_1^2 \; (\textrm{GeV}^2)$ & $100^2$ & 0 & $150^2$ & $100^2$ & 0 & $50^2$ \\
$\lambda_2$ (GeV) & $-35$ & $-35$ & $-63$ & $-35$ & $-35$ & $-63$ \\
$M_1$ (GeV) & 500 & 500 & 250 & 500 & 250 & 200 \\
$M_2$ (GeV) & 500 & 500 & 500 & 500 & 500 & 400 \\
$m_{u0}^2 \; (\textrm{GeV}^2)$ & $-100^2$ & $-150^2$ & $-150^2$ & $-100^2$ & $-150^2$ & $-150^2$ \\
$m_{d\ell}^2 \; (\textrm{GeV}^2)$ & $100^2$ & $200^2$ & $100^2$ & $100^2$ & $200^2$ & $100^2$ \\
$\mu_1^2 \; (\textrm{GeV}^2)$ & $400^2$ & $300^2$ & $300^2$ & $400^2$ & $400^2$ & $350^2$ \\
$\mu_2^2 \; (\textrm{GeV}^2)$ & $200^2$ & $300^2$ & $250^2$ & $200^2$ & $200^2$ & $300^2$ \\
$\mu_3^2 \; (\textrm{GeV}^2)$ & $200^2$ & $200^2$ & $250^2$ & $200^2$ & $250^2$ & $200^2$ \\
$\mu_4^2 \; (\textrm{GeV}^2)$ & $400^2$ & $200^2$ & $200^2$ & $400^2$ & $400^2$ & $100^2$ \\
$\mu_a$ (GeV) & 100 & 75 & 75 & 100 & 100 & 80 \\
$\mu_b$ (GeV) & 200 & 150 & 300 & 200 & 250 & 400 \\
$\mu_c$ (GeV) & 200 & 200 & 400 & 200 & 300 & 200 \\
$\mu_d$ (GeV) & 200 & 100 & 100 & 200 & 250 & 100 \\
\hline
\multicolumn{7}{|c|}{Continued on the next page} \\
\hline
\hline
\end{tabular}
\end{table}

\newpage
\addtocounter{table}{-1}

\begin{table}
\centering   \caption{continued}
\begin{tabular}{|c|c|c|c|c|c|c|}
\hline \hline
Benchmark point & I & II & III & IV & V & VI \\
\hline
$t^3 \; (\textrm{GeV}^3)$ & $60.6^3$ & $0$ & $83.9^3$ & $55.0^3$ & $0$ & $-87.9^3$ \\
$b_s^2 \; (\textrm{GeV}^2)$ & $63.4^2$ & $43.6^2$ & $98.2^2$ & $66.3^2$ & $47.1^2$ & $99.0^2$  \\
$a_s$ (GeV) & $-42.4$ & $-21.7$ & $-50.2$ & $-42.2$ & $-20.0$ & $-50.2$\\
$m_{\chi_1}$ (GeV) & 7.4 & 7.4 & 7.7 & 7.4 & 7.4 & 7.7 \\
$m_{\chi_{1}^{\pm}}$ (GeV) & 118 & 117 & 151 & 118 & 117 & 137 \\
$m_{h_1}$ (GeV) & 11.3 & 19.2 & 12.8 & 41.5 & 41.4 & 23.1 \\
$m_{a_1}$ (GeV) & 18.7 & 16.1 & 18.8 & 19.3 & 19.2 & 11.7 \\
$\langle\sigma v\rangle$  $(\tfrac{\textrm{cm}^3}{\textrm{s}})$ & $4.0 \times 10^{-26}$ & $3.4 \times 10^{-26}$ & $4.6 \times 10^{-26}$ & $3.0 \times 10^{-26}$ & $3.1 \times 10^{-26}$ & $4.1 \times 10^{-26}$  \\
$\frac{\langle\sigma v \: (\chi_1\chi_1 \rightarrow \textrm{hadrons}) \rangle}{\langle\sigma v\rangle}$ & 23\% & 38\% & 32\% & 23\% & 24\% & 30\% \\
$\sigma_{SI} (\textrm{ cm}^2)$ & $1.7 \times 10^{-40}$ & 1.2$ \times 10^{-40}$ & $1.5 \times 10^{-40}$ & $1.2 \times 10^{-42}$ & $6.1 \times 10^{-42}$ & $1.5 \times 10^{-41}$ \\
$\Gamma_{Z \rightarrow \chi_1\chi_1}$ (GeV) & $1.4 \times 10^{-9}$ & $0$  & $2.1 \times 10^{-10}$ & $1.4 \times 10^{-9}$ & $0$ & $6.3 \times 10^{-10}$ \\
$\Gamma_{Z \rightarrow h_1 a_1}$ (GeV) & $1.1 \times 10^{-11}$ & $1.2 \times 10^{-10}$ & $1.4 \times 10^{-10}$ & $4.9 \times 10^{-12}$ & $4.2 \times 10^{-11}$ & $1.2 \times 10^{-10}$ \\
$k$ & $8.0 \times 10^{-3}$ & $ 3.5\times 10^{-2}$ & $2.2 \times 10^{-2}$ & $1.3 \times 10^{-2}$ & 0.12 & $2.8 \times 10^{-2}$ \\
$S_{model} (e^+e^- \rightarrow h_1a_1)$ & $1 \times 10^{-10}$ & $2 \times 10^{-9}$ & $2 \times 10^{-9}$ & $1 \times 10^{-10}$ & $1 \times 10^{-9}$ & $2 \times 10^{-9}$ \\
$S_{model} (e^+e^- \rightarrow h_2a_1)$ & $1 \times 10^{-12}$ & $5 \times 10^{-11}$ & $3 \times 10^{-11}$ & $2 \times 10^{-12}$ & $1 \times 10^{-10}$ & $4 \times 10^{-11}$ \\
$\sigma_{e^+e^- \rightarrow \chi_1\chi_2}$ (pb) & $1 \times 10^{-5}$ & $0$ & $ 5 \times 10^{-9}$ & $1 \times 10^{-5}$ & 0 & $4 \times 10^{-6}$ \\
\hline
\hline
\end{tabular}
\end{table}

\end{document}